%% file: XS_FUZZYv6.tex
\documentclass[prl,nofootinbib,twocolumn]{revtex4}
\usepackage{epsfig,amsmath,amssymb,bm,color,graphicx}
\usepackage{afterpage}
\usepackage{pdfpages}

\def\lya{Lyman-$\alpha$\,}

\def\mFDM{m_{\rm FDM}}

\graphicspath{{./figures/}}

\newcommand{\hMpc}{h\,\mathrm{Mpc^{-1}}}
\newcommand{\skm}{\mathrm{km^{-1}\,s}}

\newcommand{\be}{\begin{equation}}
\newcommand{\ee}{\end{equation}}

\usepackage{color}

\begin{document}
\title{First constraints on fuzzy dark matter from Lyman-$\alpha$ forest data and hydrodynamical simulations}
\author{Vid Ir\v{s}i\v{c}$^{1,2,3}$}\thanks{E-mail: irsic@uw.edu (VI)}
\author{Matteo Viel$^{4,5,6}$} \thanks{E-mail: viel@sissa.it (MV)}
\author{Martin G. Haehnelt $^{7}$}
\author{James S. Bolton $^{8}$}
\author{George D. Becker$^{9,10}$}

\smallskip
\affiliation{
$^{1}$University of Washington, Department of Astronomy, 3910 15th Ave
NE, WA 98195-1580 Seattle, USA\\
$^{2}$Institute for Advanced Study, 1 Einstein Drive, NJ 08540 Princeton, USA\\
$^{3}$The Abdus Salam International Centre for Theoretical Physics,
Strada Costiera 11, I-34151 Trieste, Italy\\
$^{4}$ SISSA-International School for Advanced Studies, Via Bonomea 265, 34136 Trieste, Italy \\
$^{5}$INAF - Osservatorio Astronomico di Trieste, Via G. B. Tiepolo
11, I-34143 Trieste, Italy\\
$^{6}$INFN - National Institute for Nuclear Physics, via Valerio 2,
I-34127 Trieste, Italy\\
$^{7}$ Institute of Astronomy and Kavli Institute of Cosmology, Madingley Road, Cambridge CB3 0HA, UK \\
$^{8}$ School of Physics and Astronomy, University of Nottingham,
University Park, Nottingham, NG7 2RD, UK\\
$^{9}$ Institute of Astronomy and Kavli Institute of Cosmology, Madingley Road, Cambridge CB3 0HA, UK \\
$^{10}$Space Telescope Science Institute, 3700 San Martin Drive,
Baltimore, MD 21218, USA\\
}

\begin{abstract}

We present constraints on the masses of extremely light bosons dubbed
fuzzy dark matter from \lya forest data.  Extremely light bosons with
a De Broglie wavelength of $\sim 1$ kpc have been suggested as dark
matter candidates that may resolve some of the current small scale
problems of the cold dark matter model.  For the first time we use
hydrodynamical simulations to model the \lya flux power spectrum in
these models and compare with the observed flux power spectrum from
two different data sets: the XQ-100 and HIRES/MIKE quasar spectra
samples. After marginalization over nuisance and physical parameters
and with conservative assumptions for the thermal history of the IGM
that allow for jumps in the temperature of up to $5000\rm\,K$, XQ-100
provides a lower limit of 7.1$\times 10^{-22}$ eV, HIRES/MIKE returns
a stronger limit of 14.3$\times 10^{-22}$ eV, while the combination of
both data sets results in a limit of 20 $\times 10^{-22}$ eV
(2$\sigma$ C.L.). The limits for the analysis of the combined data
sets increases to 37.5$\times 10^{-22}$ eV (2$\sigma$ C.L.)  when a
smoother thermal history is assumed where the temperature of the IGM
evolves as a power-law in redshift.  Light boson masses in the range
$1-10 \times10^{-22}$ eV are ruled out at high significance by our
analysis, casting strong doubts that FDM helps solve the "small
  scale crisis" of the cold dark matter models.

\end{abstract}

\maketitle {\it Introduction.}  Recently, there has been a growing
interest in so-called Fuzzy Dark Matter (FDM) models where the dark
matter is made of ultra-light bosons. Cosmological and astrophysical
consequences have been comprehensively reviewed in
\citep{hu2000,matos00,hlozek,hui16,bernal17} highlighting the particle
physics
motivation \citep{baldeschi83,preskill83,sikivie,zhang17} for such
models, as well as the
importance of experimental searches \citep{sikivie83}. A broad variety
of astrophysical implications have been investigated in the
literature: the halo mass function \citep{marsh}, the innermost
structure of haloes \citep{lee96,guzman99,zhang16}, the dynamical
properties of the
smallest objects \citep{calabrese}, the linear matter power spectrum
\citep{hu2000}, the development of non-linearities by using N-body
simulations \citep{schive14}, the abundance of high redshift objects
\citep{menci17}, the overall impact of FDM on galaxy formation and the
reionization history of the Universe, the intergalactic medium
\citep{amendola06,kim16,marsh16,hui16,sarkar17}, pulsar timing and
binary
pulsars \citep{blas16,pulsarfuzzy}, and the properties of our galactic
disk \citep{banik17}. The general conclusion is that in order to have
an appreciable astrophysical impact the mass of ultra-light bosons
would have to lie in the range $1-10 \times 10^{-22}$ eV, and in this
mass range it is indeed possible that some small scale ``tensions'' of
cold dark matter with observations could be alleviated
(e.g. \citep{weinberg15} for a review of the small scale ``crisis'' of
cold dark matter).

The intergalactic medium (IGM) \citep{meiksin09,mcquinn15} plays a
unique role in constraining the (small scale) matter power spectrum,
since the low-density, high redshift IGM filaments are particularly
sensitive to the small scale properties of dark matter.  The main
observable manifestation of the IGM, the Lyman-$\alpha$ forest, has
provided important constraints on the linear matter power spectrum,
especially when combined with cosmic microwave background
data \citep{croft02,zaldarriaga03,mcdonald03,viel04,mcdonald05,seljak06,lidz10}. This
includes, most notably, the tightest constraints on warm dark matter
(WDM) models \citep{viel05,viel13WDM,uros06,baur15,viel08,yeche17}, upper
limits on neutrino masses \citep{seljak06,palanque15,yeche17} as well as the
recent remarkable discovery of Baryonic Acoustic Oscillations in the
transmitted 3D flux \citep{busca13,slosar13}. These results,
especially those at small scales, are primarily due to the fact that
the observed Lyman-$\alpha$ forest flux power spectrum provides a
tracer of matter fluctuations on small scales and at high redshifts,
where these fluctuations are still in the quasi-linear regime.
At present, the tightest limits on the free streaming of
WDM, expressed as the equivalent masses of thermal WDM relics, are in
the range m$_{\mathrm{WDM}}>2.3-5$ keV at 2$\sigma$ C.L.. The values
at the lower end of this range are probably overly conservative and
require the assumption of thermal histories that are likely unphysical
\citep{garzilli15}. 

In FDM models, even though the typical de Broglie scale is rather small
($\sim$ kpc) the effect of FDM on the linear matter power spectrum is
noticeable on scales larger than the smallest scales typically
constrained by IGM data \citep{hu2000}.  In the absence of fully
numerical FDM simulations of the flux power spectrum, it has therefore
become common practice to convert the limit on thermal relic WDM
models -- for which the flux power spectrum has been modelled in
considerable detail -- into FDM limits by comparing the linear matter
power spectrum of WDM and FDM models and using the mass corresponding
to $k_{\rm 1/2}$, the wavenumber at which the linear power spectrum
departs (i.e. is suppressed) from the corresponding cold dark matter
power spectrum by 50\%.  However, the accuracy of this rather crude
mapping can only be checked by performing a full set of hydrodynamical
simulations to model the effect of FDM on the properties of the IGM
and the \lya forest.  Here, we will use such simulations to provide the
first constraints on FDM models based on a full modelling of the \lya
flux power spectrum and comparison with two high-redshift data sets
well suited to probing the small scale matter power spectrum. This
will also allow us to check the accuracy of the $k_{\mathrm 1/2}$
mapping of thermal relic WDM constraints.  Our analysis will be quite
similar to the one presented in \citep{viel13WDM} and \citep{irsic17},
where the flux power spectrum is modelled using a set of
hydrodynamical simulations that vary astrophysical and cosmological
parameters combined with a Monte Carlo Markov Chain analysis in the
multi dimensional parameter space.

{\it Data.}  The first sample we use is the set of 100 medium resolution,
high signal-to-noise QSO spectra of the XQ-100 survey \citep{lopez16}, with
emission redshifts $3.5 < z < 4.5$.  A more detailed description of
the data and the power spectrum measurements of the XQ-100 survey can
be found in \citep{irsic17}.  Here we repeat the most important
properties of the data and the derived flux power spectrum.  The
spectral resolution of the X-shooter spectrograph is 30-50 km/s,
depending on wavelength.  The flux power spectrum used in the analysis
has been calculated for a total of 114 $(k,z)$ data points in the
ranges $z=3,3.2,3.4,3.6,3.8,4,4.2$ and 19 bins in $k-$space in the
range 0.003-0.057 s/km.  We further use the measurements of the flux
power spectrum presented in \citep{viel13WDM}, at redshift bins
$z=4.2,4.6,5.0,5.4$ and in 10 $k-$bins in the range 0.001-0.08 s/km.
In this second sample the spectral resolution of the QSO absorption
spectra obtained with the MIKE and HIRES spectrographs are about 13.6
and 6.7 km/s, respectively.  As in the analysis of \citep{viel13WDM},
a conservative cut is imposed on the flux power spectrum obtained from
the MIKE and HIRES data, and only the measurements with $k >
0.005\;\skm$ are used to avoid possible systematic uncertainties on
large scales due to continuum fitting.

Compared to XQ-100, the HIRES/MIKE sample has the advantage of probing
smaller scales and higher redshift.  There is a small redshift overlap
between the two samples at $z=4.2$. Since the thermal broadening
(measured in km/s) of Lyman-$\alpha$ forest lines is roughly constant
with redshift, the presence of a cutoff in the matter power spectrum
due to free-streaming becomes more prominent in velocity space at high
redshift due to the $H(z)/(1+z))$ scaling between the fixed comoving
length scale set by the free-streaming length and the corresponding
velocity scale.  Moreover, the 1D power spectrum is more sensitive to
the presence of a cutoff compared to the 3D power spectrum.

\begin{figure}
\begin{center}
\includegraphics[width=8.5cm,height=8cm]{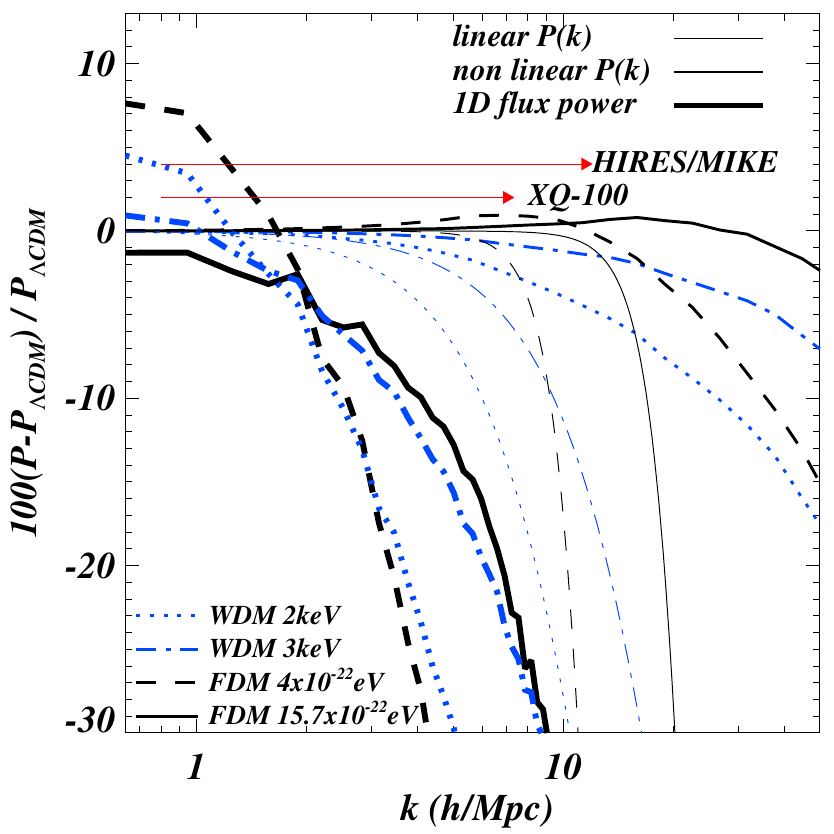}
\caption{Power spectrum relative to $\Lambda$CDM at $z=5.4$ (in per
  cent). Linear matter, non-linear matter and flux power spectra are
  represented by the thin, thick and very thick curves,
  respectively. Black (blue) curves are for FDM (WDM) models with
  m$_{\rm FDM}=5.7, 15.7 \times 10^{-22}$ eV (m$_{\rm WDM}$ = 2, 3
  keV).
\label{fig_0}}
\end{center}
\end{figure}

{\it Simulations.} Similarly as in \citep{viel13WDM} and
\citep{irsic17}, we model the flux power spectrum using a set of
hydrodynamical simulations performed with the GADGET-3 code, a
modified version of the publicly available Gadget-2 code
\citep{gadget}.  A simplified star formation criterion is applied for
which gas particles above an overdensity 1000, and below T$=10^5$ K
are converted into stars (e.g. \citep{bolton16}). The reference model
simulation has a box length of 20 Mpc$/h$ with $2 \times 768^3$ and
(cold) dark matter particles (gravitational softening of 1
comov. kpc$/h$) in a flat $\Lambda$CDM Universe with cosmological
parameters $\Omega_{\rm m} = 0.301$, $\Omega_{\rm b} = 0.0457$,
$n_{\rm s} = 0.961$, $H_0 = 70.2\;\mathrm{km\,s^{-1}\,Mpc^{-1}}$ and
$\sigma_8 = 0.829$ in agreement with \citep{planck15}. Three different
WDM models with masses $m_{\rm WDM} = 2, 3, 4\;\mathrm{keV}$ have also
been simulated in order to obtain WDM constraints.

We simulate 5 different FDM models using the transfer function
provided by \citep{hu2000} with light axion masses m$_{\rm FDM}$ of 1,
4, 5.7, 15.7 and 30 $\times 10^{-22}$ eV.  These values have roughly the same
k$_{1/2}$ as models
corresponding  to thermal WDM relic masses of 1, 1.73, 2, 3, 3.87 keV,
covering the range relevant for the ``small scale crisis'' of cold dark matter. 
These models were also simulated using the axionCAMB code \citep{hlozek} to obtain
the linear transfer function, finding negligible impact on the simulated flux power.
The corresponding $\Lambda$CDM model is also
simulated along with a range of IGM thermal histories and cosmological
parameters. In Fig.~\ref{fig_0} we show the linear, non-linear and
flux power spectrum at $z=5.4$ for WDM and FDM models that have the
same k$_{1/2}$: 
non-linearities erase some of the information contained in
the linear power spectrum. Note that the 1D flux power is much more
sensitive to the cutoff. The maximum wavenumbers at which the flux power
spectrum is measured by HIRES/MIKE and XQ-100 are represented by the
horizontal arrows. 

It has been noted before (e.g. \citep{schive16}) that
for the analysis of the \lya\ it is sufficient to use the appropriate transfer
function without modelling the full quantum effects below the
de-Broglie wavelength of the FDM particle. This hypothesis is
supported by the fact that the quantum pressure starts to dominate
over gravity on scales smaller than the FDM Jeans scale ($k > k_J$)
\citep{schive16}. The FDM Jeans scale increases with cosmic time, and
also increases with the mass of the FDM particle. For the largest redshift,
$z=5.4$, and smallest mass in our simulations, $m_{\rm FDM} = 1 \times
10^{-22}\;\mathrm{eV}$, the FDM Jeans scale is $64.7\,\hMpc$, which corresponds to 
scales smaller than the scale probed by our data ($k_{\mathrm{max}} =
12.7\,\hMpc$, for HIRES at $z=5.4$). \footnote{Furthermore, the growth rate
ratio $\xi$, as defined in \citep{schive16}, is $\xi > 0.995$ for
scales $k < 12.7\,\hMpc$ for the redshift range considered in this
paper. The value of $\xi$ decreases as we approach the FDM Jeans
scale, with $\xi = 0.55$ at $k = 40\,\hMpc$.} The effect of quantum
pressure term should thus have a negligible effect on the structure
formation relevant for the \lya\ forest.

We vary the thermal history by modifying the photo-heating rates in
the simulations as in \citep{bolton08}.  The low density IGM
($\Delta=1+\delta < 10$) is well described by a power-law
temperature-density relation, $T = T_0\Delta^{\gamma-1}$.  We consider
a range of values for the temperature at mean density $T_0$ and the
slope of the $T-\rho$ relation, $\gamma$, based on the previous
analysis of the \lya\ forest and recent observations \citep{becker11}.
As in \citep{irsic17} these consist of a set of three different
temperatures at mean density, $T_0(z=3.6) = 7200, 11000,
14800\;\mathrm{K}$, which evolve with redshift, as well as a set of
three values of the slope of the $T-\rho$ relation: $\gamma(z=3.6) =
1.0, 1.3, 1.5$.  The reference thermal history assumes
$(T_0(z=3.6),\gamma(z=3.6))= (11000\;\mathrm{K},1.5)$.
  
Following again \citep{irsic17} we use two parameters describing
cosmology, $\sigma_8$ and $n_{\rm eff} = d\ln{P_{\rm m}(k)}/d\ln{k}$,
evaluated at $k = 0.005\;\skm$. Five different values are considered
for both $\sigma_8 = 0.754,0.804,0.829,0.854,0.904$, and $n_{\rm eff}
= -2.3474, -2.3274,-2.3074,-2.2874, -2.2674$. The reference model has
$(\sigma_8,n_{\rm eff}, n_{\rm s})= (0.829,-2.3074,0.961)$.  We also
vary the redshift of reionization $z_{\rm rei}$ which is chosen to be
$z_{\rm rei} = 9$ for the reference model as well as $z_{\rm rei} =
7,15$ for two additional models \citep{supplemental}. The last parameter ($f_{\rm UV}$) characterizes the
effect of Ultraviolet (UV) background fluctuations. An extreme model
dominated by QSOs has been chosen with a strong scale dependence at
higher redshift and towards large scales. The mean flux is also
varied a-posteriori through rescaling the effective optical depth,
$\tau_{\rm eff}=-\ln {\bar F}$. We use three different values
$(0.8,1,1.2) \times \tau_{\rm obs, eff}$, with the observed value of
$\tau_{\rm obs,eff}$ chosen to be those of the SDSS-III/BOSS
measurements \citep{palanque13}.

{\it Method.}  Using the models of the flux obtained from the
simulations we establish a grid of points for each redshift, in the
parameter space of $({\bar F}(z), T_0(z), \gamma(z), \sigma_8, z_{\rm
  rei}, n_{\rm eff}, f_{\rm UV}, m_{\rm FDM})$.  We then perform a
linear interpolation between the grid points in this multidimensional
parameter space, to obtain predictions of flux power for the desired
models.  The interpolation is performed for $P_{\rm F}(k,z)$,
directly, rather than for ratios of flux power w.r.t. the
corresponding $ \Lambda CDM$ simulation as was done in
\cite{viel13WDM}.  Parameter constraints are then obtained with a
Monte Carlo Markov Chain code that explores the likelihood space until
convergence is reached.

\begin{figure}
\begin{center}
\includegraphics[width=9cm]{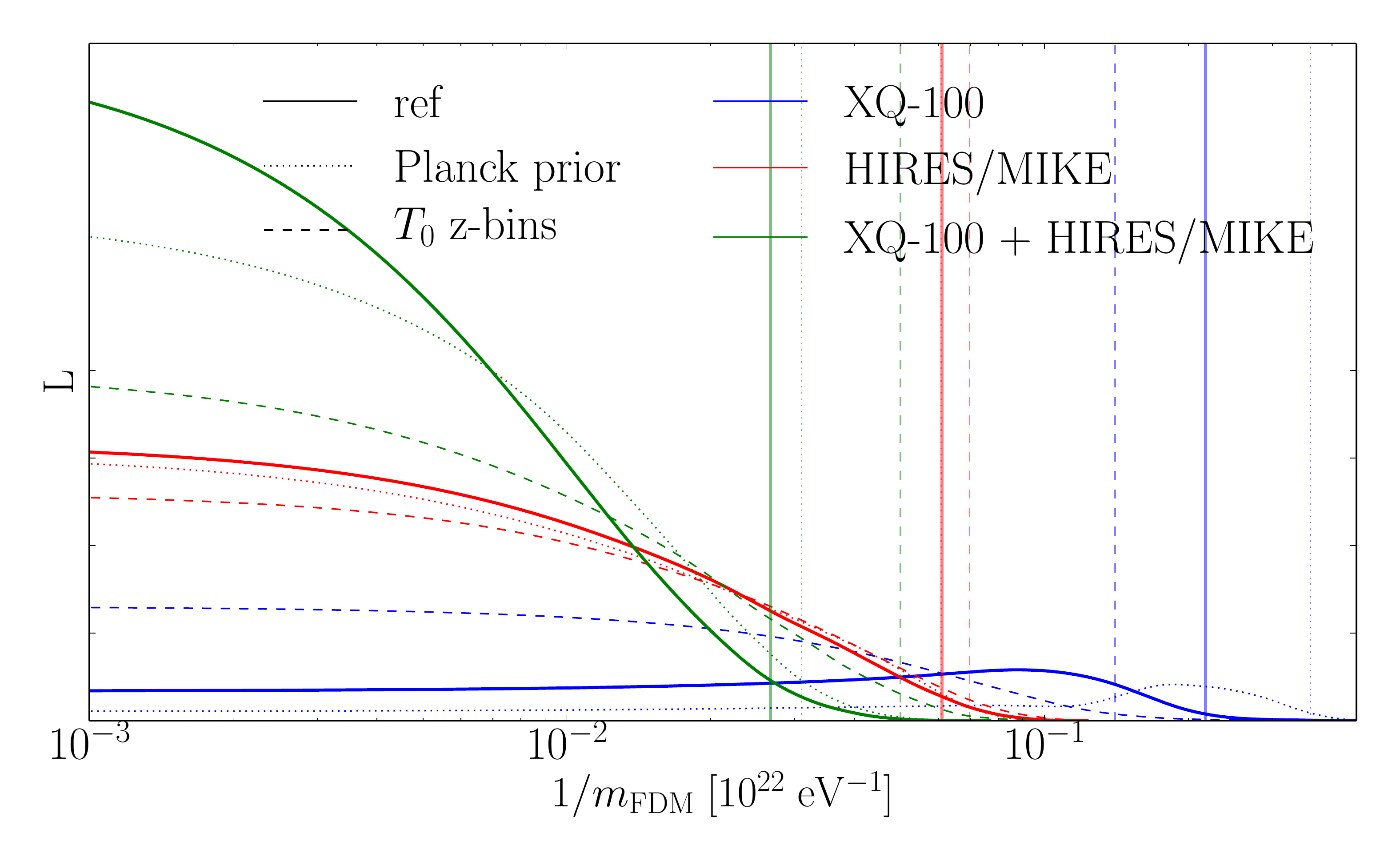}
\caption{1D marginalized likelihood constraints for $1/m_{\rm
    {FDM}}$. XQ-100, HIRES/MIKE and XQ-100+HIRES/MIKE are represented
  by the blue, red and green curves, respectively, with (continuous
  curves) and without (dotted curves) using Planck
  priors. Additionally results for models where we vary temperature
  independently in each redshift bin are plotted as dashed curves. 2$\sigma$
  upper limits are represented by vertical lines.
\label{fig_1}}
\end{center}
\end{figure}

The redshift evolution of the IGM parameters $T_0$ and $\gamma$ are
modelled as power-laws for our reference analysis: $T_0(z) =
T_0^A\left[(1+z)/(1+z_{\rm p})\right]^{T_0^S}$ and $\gamma(z) =
\gamma^A\left[(1+z)/(1+z_{\rm p})\right]^{\gamma^S}$. The pivot
redshift is different for each data set and roughly corresponds to the
redshift at which most of the data lies
($z_{\rm p} = 3.6, 4.5, 4.2$ for XQ-100, HIRES/MIKE and the combined
analysis, respectively).

{\it Results.}  In Figure~\ref{fig_1} we show the main result of this
letter: the marginalized 1D likelihood for 1/m$_{\rm FDM}$. For our
reference analysis, in which the temperature evolution is
parameterized as a power-law at different pivot redshifts, XQ-100
returns an upper limit of $4.6 \times 10^{-22}$ eV, HIRES/MIKE gives
16.4$\times 10^{-22}$ eV, while the combination of the two data sets
results in a considerable improvement to 37.5$\times 10^{-22}$ eV
(2$\sigma$ C.L.). These numbers become 2.7, 16.5, 32.2 $\times
10^{-22}$ eV, for XQ-100, HIRES/MIKE and both data sets, when using
the following Planck priors on n$_{\rm eff} = -2.307 \pm 0.01$ and
$\sigma_8 = 0.829 \pm 0.01$ (1$\sigma$ Gaussian priors).  The
improvement in the joint constraints is due to the fact that when
combining the two data sets the thermal evolution of the IGM is assumed to be one
power-law for both $T_0(z)$ and $\gamma(z)$ for the full redshift range of the combined data sets. 
If we
drop the assumption of a power-law evolution and we let the
temperature vary independently in each redshift bin, with a maximum
jump $\Delta T= 5000\,$K in bins that are separated by $\Delta z=0.2$, we
obtain 7.1, 14.3, 20.0 $\times 10^{-22}$ eV for XQ-100, HIRES/MIKE and
both combined. We regard this result as the most conservative, since
sudden jumps of temperature are not physically plausible in this
redshift range (e.g. \citep{puchwein15,upton}).
  
 Increasing the covariance matrix by a multiplicative factor 1.3 in
 order to better represent a possible underestimation of the errors
 does not affect the results appreciably. In Tab.~1 we summarize the
 results including the $\chi^2/d.o.f.$ for the reference case, which
 appear to be very reasonable in all cases. For the combined analysis
 the other parameters lie within the following 2$\sigma$ C.L. ranges
 at $z_{\rm p}=4.2$: $\sigma_8=[0.83,0.95]$, $n_{\rm
   eff}=[-2.43,-2.31]$, $T^A(z_{\rm
   p})\;\mathrm{[10^4\,K]}=[0.71,1.06]$, $T^S(z_{\rm
   p})=[-3.37,-0.80]$, $\gamma^A(z_{\rm p})=[1.27,1.69]$,
 $\gamma^S(z_{\rm p})=[-0.11,1.82]$, $z_{\rm rei}=[6.27,13.62]$,
 $f_{\rm UV}=[0.04,0.94]$.

We have verified the constraints obtained by considering additional
simulations with $m_{\rm FDM}$ of 30 $\times 10^{-22}$ eV.

\begin{table}[h]
\small
\begin{tabular}{lccc}
\hline
$\mFDM\;\mathrm{[10^{-22} eV]} $ & {\rm XQ-100} & {\rm HIRES/MIKE} & {\rm Combined}\\
\hline
ref. & $4.5$& $16.4$ & $37.5$ \\
Cov. $\times$ 1.3 & $3.9$& $16.3$ & $34.9$ \\
Planck priors  & $2.7$& $16.5$ & $32.2$ \\
$T_0(z)$ bins & $7.1$& $14.3$ & $20.0$ \\
\hline
$\chi^2/d.o.f.$ (ref.) & $134/124$& $33/40$& $187/173$\\
\hline
\end{tabular}
\caption{Marginalized constraints at $95$ \% (lower limits). The pivot
  redshifts for different data sets are: $z_{\rm p}=3.6, 4.5, 4.2$ for
  XQ-100, HIRES/MIKE and combined for reference case, covariance
  matrix multiplied by 1.3, Planck priors and temperature in redshift
  bins.}
\label{tab:constraints}
\end{table}

\begin{figure}
\begin{center}
\includegraphics[width=9cm]{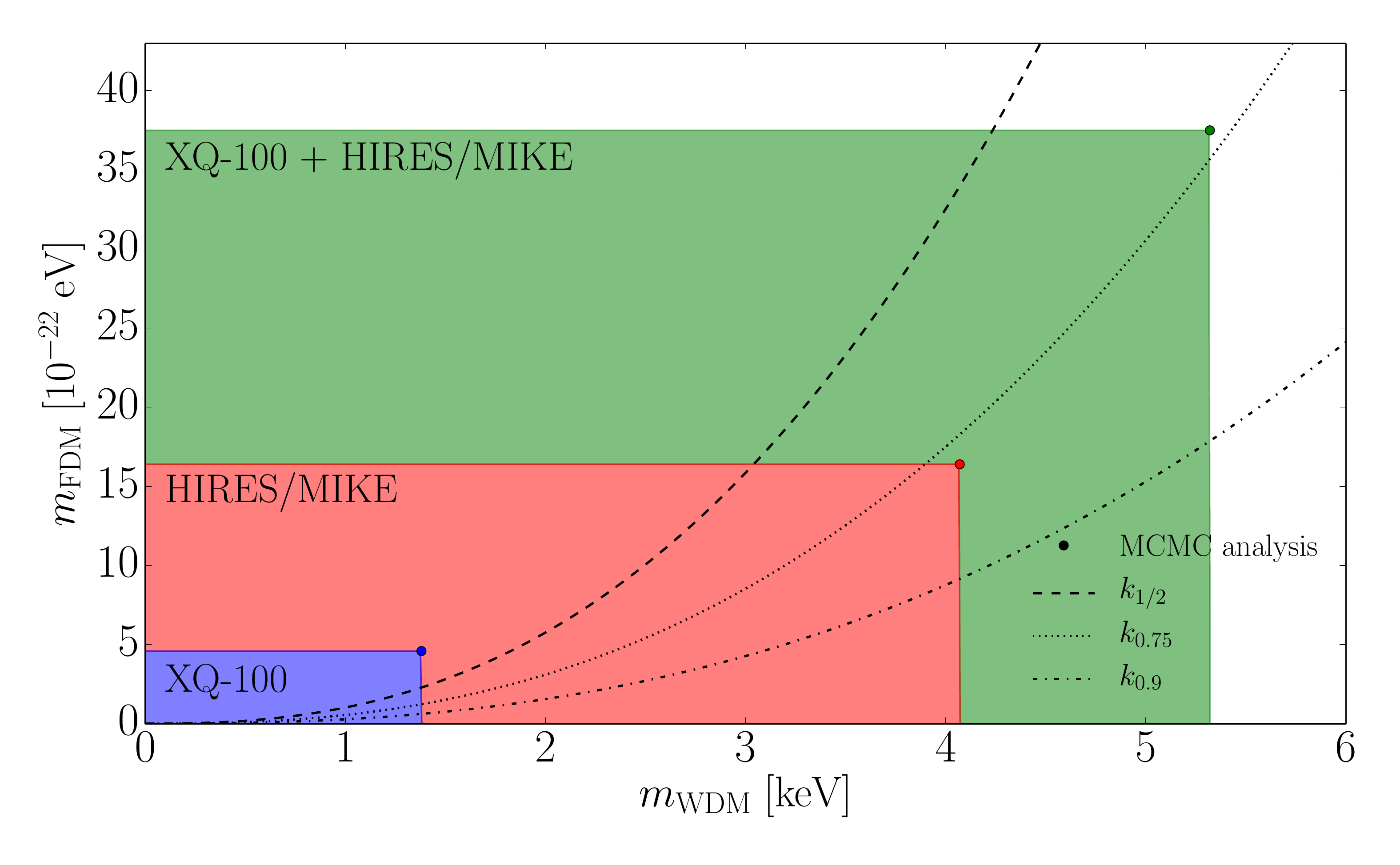}
\caption{Comparison of limits on $m_{\mathrm {FDM}}$ and $m_{\rm WDM}$
  (thermal relic). Also shown is the relation obtained by matching
  $k_{\mathrm{1/2}}$, $k_{\mathrm{0.75}}$ and $k_{\mathrm{0.9}}$ for
  the linear power spectra. Shaded areas show regions excluded by the
  analysis of XQ-100, HIRES/MIKE and both data sets in blue, red and
  green colors, respectively. Filled circles represent the 2$\sigma$
  C.L. lower limits for WDM and FDM.\label{fig_2}}
\end{center}
\end{figure}

 In Figure~\ref{fig_2} we illustrate the relation between WDM and FDM
 constraints.  Due to the lack of the detailed modelling of the flux
 power spectrum for FDM models using hydrodynamical simulation so far,
 it has been usually assumed that one can map thermal relic WDM masses
 into FDM constraints by identifying the corresponding k$_{1/2}$
 values for the linear power spectra, i.e. the wavenumber at which the
 power reaches 50\% of the $\Lambda$CDM linear power spectrum.
 Figure~\ref{fig_2} shows that this is not a good approximation.  For
 XQ-100 only, i.e. for a WDM constraint of 1.34 keV, the corresponding
 FDM limit is slightly above the k$_{1/2}$ curve, which remains a
 reasonable approximation. The HIRES/MIKE data only, however, gives a
 WDM lower limit of 4.7 keV that translates into a FDM limit which is
 much weaker than the one that would be obtained using the k$_{1/2}$
 mapping.  The same holds for the combined analysis.  For these
 higher thermal relic WDM masses the WDM constraints are better mapped
 into FDM constrains by using k$_{0.75}$ rather than k$_{1/2}$, where
 k$_{0.75}$ is defined as the wavenumber at which the power reaches
 75\% of the $\Lambda$CDM linear power spectrum.  The reason for this
 is that as the free-streaming cut-off moves to smaller scales for
 larger particle masses, the scales affected by the free-streaming
 cut-off become more non-linear.

 We have also checked that similar conclusions are reached when
 considering the 1D linear power spectrum, which is the quantity
 really probed by the flux power spectrum (rather than the 3D power
 spectrum). The different data sets are obviously constraining linear
 power at different $(k,z)$ values.  Non-linearities will develop
 differently in WDM and FDM models and this makes the mapping between
 the two scenarios not straightforward.  Full non-linear hydrodynamic
 simulations and detailed modelling of the flux power spectrum is
 required for accurate constraints on FDM models.  WDM models retain
 more small scale power at $k>k_{0.75}$ compared to the corresponding
 FDM modelq that has instead a more prominent knee at $k<k_{0.75}$.
 These differences partially compensate in terms of non linear 1D flux
 power in a non-trivial way that depends on how non-linear the matter
 power spectrum is at the free-streaming scale. Similar to the shape
 of the WDM cutoff, the FDM cutoff in the flux power spectrum appears
 to be rather distinctive with no significant degeneracies with the
 other parameters in the analysis.

{\it Conclusions.}  We have presented constraints on FDM models based
on detailed modelling of the \lya forest 1D flux power spectrum and
high resolution data at intermediate and high redshifts with
hydrodynamical simulations. These are the first constraints that
incorporate the effect of the relevant IGM physics, including thermal
and pressure smoothing on the non-linear evolution of the flux power
spectrum on the relevant scales. Our final, conservative lower limit
from a joint analysis of the intermediate and high-resolution \lya
forest data, m$_{\mathrm {FDM}} > 20 \times 10^{-22}$ eV (2$\sigma$
C.L.), was obtained with conservative assumptions for the thermal
history of the IGM that allow for (unphysical) sudden jumps of the IGM
temperature up to 5000K.  This lower limit for the mass of ultra-light
bosons strengthens by about a further factor two if we assume a
smoother thermal history of the IGM.  Our analysis appears to close
the window of FDM models with significant astrophysical implications,
in particular for alleviating the tension between observations and
theoretical predictions of cold dark matter models on small scales.
{\footnote {After this paper was completed an analysis of IGM data has been performed
by \cite{armengaud17} reaching similar conclusions with slightly weaker constraints
on FDM mass.}}

\begin{acknowledgments}
VI is supported by US NSF grant AST-1514734. VI also thanks M. McQuinn
for useful discussions, and IAS, Princeton, for hospitality during his
stay where part of this work was completed. MV is supported by
INFN/PD51 Indark and by the ERC Grant 257670-cosmoIGM and by PRIN-INAF
"2012 "The X-Shooter sample of 100 quasar spectra at $z \sim
3.5$". JSB is supported by a Royal Society URF. MGH is supported by
the FP7 ERC Grant Emergence-320596 and the Kavli
Foundation. GB is supported by the NSF
under award AST-1615814. Simulations were performed at the University of Cambridge
with Darwin-HPCS and COSMOS, operated on behalf of the STFC DiRAC
facility (funded by BIS National E-infrastructure capital grant
ST/J005673/1 and STFC grants ST/H008586/1, ST/K00333X/1).
\end{acknowledgments}

\input{supplemental}

\bibliographystyle{unsrtnat}
\bibliography{Bibliofile}

\end{document}

%% file: supplemental.tex






\section*{Supplemental material}
\subsection*{The effect of pressure smoothing}

The gas physics imprints two distinct scales on the distribution of
the flux. The first is  the thermal Doppler broadening, which smoothes the
optical depth along the line of sight and this effect depends on the
temperature at a given redshift. The second effect is from Jeans
thermal pressure smoothing which affects the gas distribution relative
to the 3D dark matter distribution. Because it takes some time for the
gas to respond to the changes in temperature, this effect depends on
the thermal history evolution at earlier times.

Similarly to pressure smoothing, additional smoothing due to FDM model
also affects the 3D distribution. Thus a certain degree of degeneracy
is expected.

In the analysis presented in this paper we have parametrised the
effects of thermal physics by three main observables ($T_0$,$\gamma$
and $z_{\rm rei}$), with further parameterisation of the redshift
evolution of $T_0$ and $\gamma$ (see main text for details). The
pressure smoothing scale would thus  mostly differ from the Doppler
broadening scale due to the value of the redshift of reionisation
($z_{\rm rei}$).

\begin{figure}[!h]
\begin{center}
\includegraphics[width=9cm,height=4.5cm]{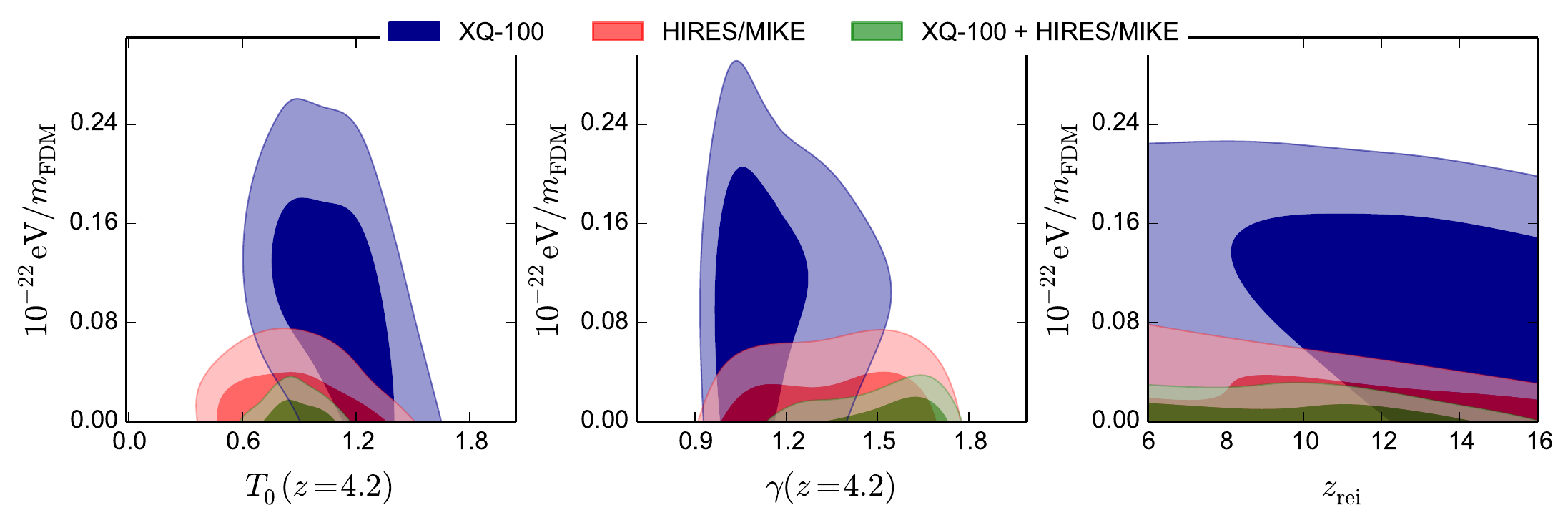}
\caption{The 2D contours between the thermal parameters and the FDM
mass. Different colour schemes represent analysis obtained using
XQ-100 only data set (blue contours), HIRES/MIKE (red contours) and
finally XQ-100 and HIRES/MIKE data sets combined (green contours).
\label{fig_s2}}
\end{center}
\end{figure}

Due to the different manifestation of the smoothing (1D vs 3D) between the
thermal parameters, as well as different redshift evolution between
thermal smoothing scales and FDM smoothing scale, degeneracies in the
parameter space can be broken. This is shown in 
Fig.~\ref{fig_s2}. The left hand panel shows that the temperature
($T_0$ at a chosen redshift) is degenerate with the FDM particle mass,
particularly when estimating the parameters in the lower redshift range
(blue contours -  XQ-100 data set). The degeneracy is broken
efficiently when moving to higher redshift range (red contours -
HIRES/MIKE data) and almost gone when including all the data (green
contours).

Moreover, the right hand side of Fig.~\ref{fig_s2} shows the
degeneracy between the $m_{\rm FDM}$ and $z_{\rm rei}$ parameters. It
is clear that when performing the analysis on higher redshift data,
and increasing the redshift range, the degeneracy between these two
parameters is almost broken. This indicates the importance of the
different redshift evolution of the smoothing scales, for estimating
the parameter values.

To further support this, we have estimated the thermal Jeans pressure
scale following
\citep{gnedin98}. 
As was shown in \citep{gnedin98}, the pressure
smoothing scale (in $\mathrm{Mpc}$/h) will always be smaller than the
instantaneous Jeans scale. Moreover, a more recent analysis on the
simulations has shown that the true pressure smoothing scale (at
redshift concerning the \lya\ forest) lies
somewhere between the filtering and Jeans smoothing
scale \citep{kulkarni15}. 

\begin{figure}[!h]
\begin{center}
\includegraphics[width=9cm]{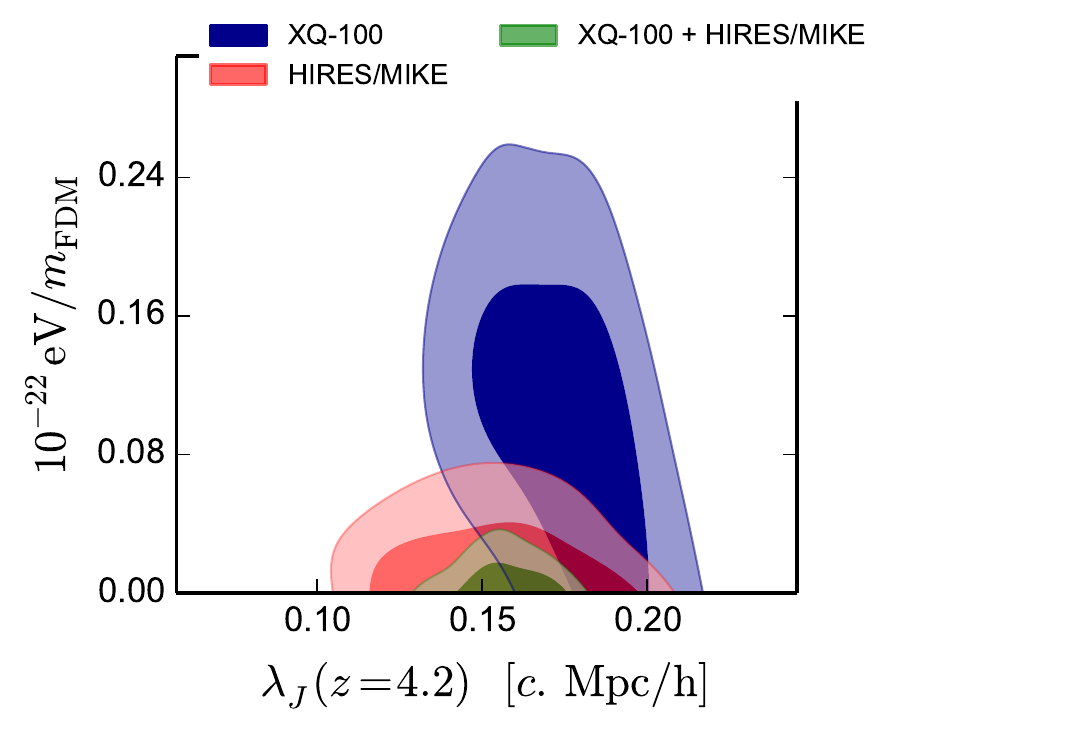}
\caption{The figure shows the 2D contours between $m_{\rm FDM}$
(governing FDM model smoothing) and the Jeans smoothing scale. 
The colour schemes reflect different data sets used in the
analysis - XQ-100 only (blue), HIRES/MIKE (red) and combined data sets
(green). The extent of smoothing increases with increasing values of
both axis, i.e. larger value of $1/m_{\rm FDM}$ implies more smoothing
due to FDM model, while larger value of $\lambda_J$
implies more thermal smoothing.
\label{fig_s3}}
\end{center}
\end{figure}

As was already
discussed, the degeneracy between $z_{\rm rei}$ and $m_{\rm FDM}$ is
very small, thus the expected difference between degeneracies in
$m_{\rm FDM}$ vs $\lambda_J$ plane will be small. This
is illustrated in Fig.~\ref{fig_s3}. The figure further shows (in the
plane of the smoothing scales) that while degeneracy exists - either
there is more thermal smoothing and less FDM smoothing or viceversa -
it is severely broken when using higher redshift data, and when
extending the redshift range of the analysis considered.

\newpage
\subsection*{The best-fit FDM model}

\begin{figure}[!h]
\begin{center}
\includegraphics[width=9.5cm]{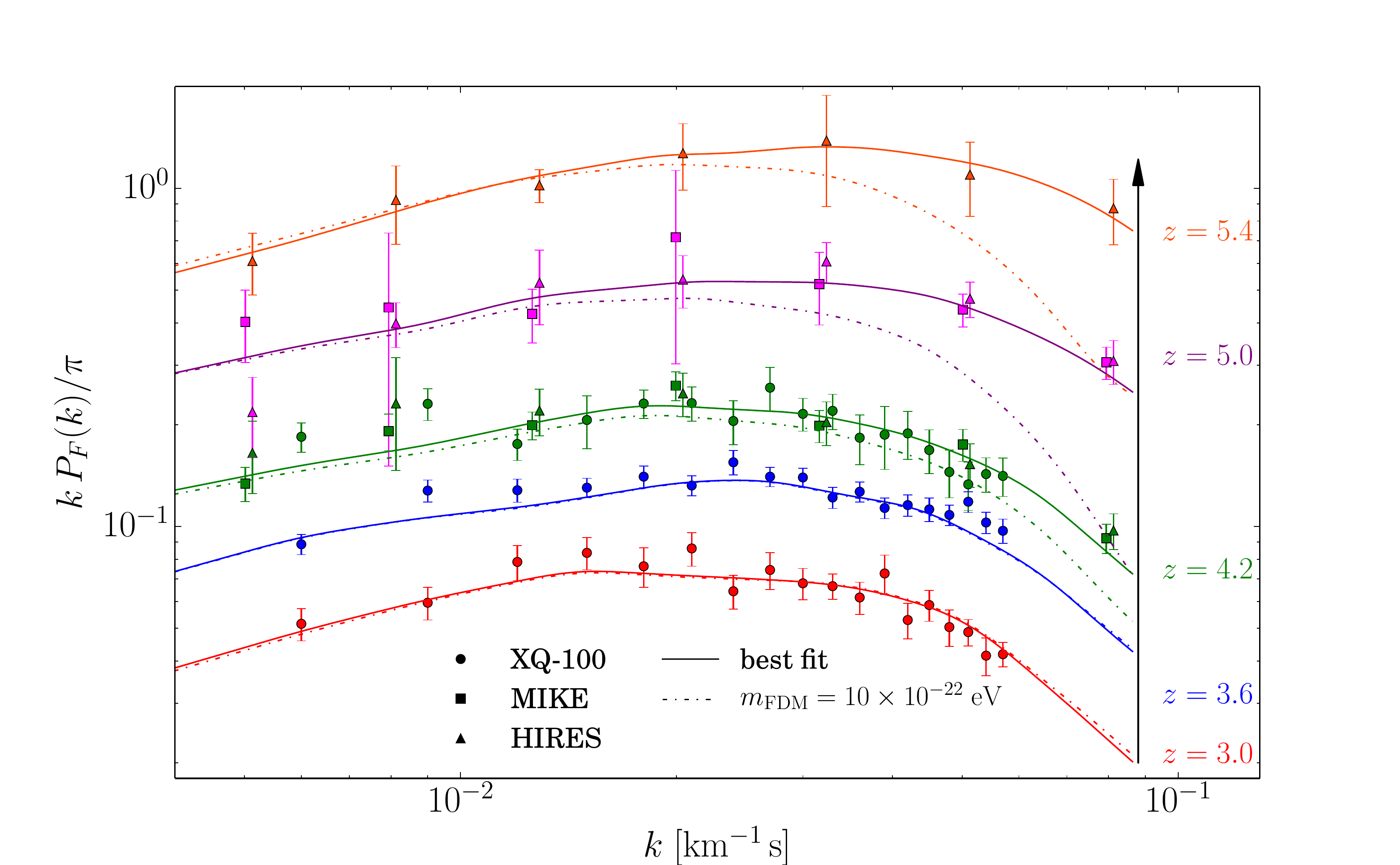}
\caption{The figure shows best-fit flux power spectrum obtained with
analysis on the combined data sets (XQ-100 - circles, MIKE - squares
and HIRES - triangles) using the REF prior model (see main text for
detail). Different colours represent different redshifts, that are
indicated on the right hand side of the plot, ranging from $z=3$ to
$z=5.4$. The dashed lines show the flux power spectrum when all the
parameters were kept at their best-fit values, except the mass of the
FDM particle was decreased to $10\;\times\mathrm{10^{-22}eV}$.
\label{fig_s4}}
\end{center}
\end{figure}

The Fig.~\ref{fig_s4} shows the data sets used in the analysis along
with the best-fit model of the power-spectrum, when performing the
analysis on the combined data set using the REF prior model. For comparison the plot
also shows the power spectrum where the parameters were kept fixed at
best-fit values, except the FDM particle mass was decreased down to
$10\times\mathrm{10^{-22}eV}$ to guide the eye. The plot nicely
illustrates that the shape of the cutoff is the most constraining effect when
considering FDM models.

%% file: XS_FUZZYv6.bbl
\begin{thebibliography}{59}
\providecommand{\natexlab}[1]{#1}
\providecommand{\url}[1]{\texttt{#1}}
\expandafter\ifx\csname urlstyle\endcsname\relax
  \providecommand{\doi}[1]{doi: #1}\else
  \providecommand{\doi}{doi: \begingroup \urlstyle{rm}\Url}\fi

\bibitem[{Hu} et~al.(2000){Hu}, {Barkana}, and {Gruzinov}]{hu2000}
W.~{Hu}, R.~{Barkana}, and A.~{Gruzinov}.
\newblock {Fuzzy Cold Dark Matter: The Wave Properties of Ultralight
  Particles}.
\newblock \emph{Physical Review Letters}, 85:\penalty0 1158--1161, August 2000.
\newblock \doi{10.1103/PhysRevLett.85.1158}.

\bibitem[{Matos} et~al.(2000){Matos}, {Guzm{\'a}n}, and
  {Ure{\~n}a-L{\'o}pez}]{matos00}
T.~{Matos}, F.~S. {Guzm{\'a}n}, and L.~A. {Ure{\~n}a-L{\'o}pez}.
\newblock {Scalar field as dark matter in the universe}.
\newblock \emph{Classical and Quantum Gravity}, 17:\penalty0 1707--1712, April
  2000.
\newblock \doi{10.1088/0264-9381/17/7/309}.

\bibitem[{Hlozek} et~al.(2015){Hlozek}, {Grin}, {Marsh}, and
  {Ferreira}]{hlozek}
R.~{Hlozek}, D.~{Grin}, D.~J.~E. {Marsh}, and P.~G. {Ferreira}.
\newblock {A search for ultralight axions using precision cosmological data}.
\newblock \emph{\prd}, 91\penalty0 (10):\penalty0 103512, May 2015.
\newblock \doi{10.1103/PhysRevD.91.103512}.

\bibitem[{Hui} et~al.(2016){Hui}, {Ostriker}, {Tremaine}, and {Witten}]{hui16}
L.~{Hui}, J.~P. {Ostriker}, S.~{Tremaine}, and E.~{Witten}.
\newblock {On the hypothesis that cosmological dark matter is composed of
  ultra-light bosons}.
\newblock \emph{ArXiv e-prints}, October 2016.

\bibitem[{Bernal} et~al.(2017){Bernal}, {Fern{\'a}ndez-Hern{\'a}ndez}, {Matos},
  and {Rodr{\'{\i}}guez-Meza}]{bernal17}
T.~{Bernal}, L.~M. {Fern{\'a}ndez-Hern{\'a}ndez}, T.~{Matos}, and M.~A.
  {Rodr{\'{\i}}guez-Meza}.
\newblock {Rotation Curves of High-Resolution LSB and SPARC Galaxies in Wave
  (Fuzzy) and Multistate (Ultra-light Boson) Scalar Field Dark Matter}.
\newblock \emph{ArXiv e-prints}, January 2017.

\bibitem[{Baldeschi} et~al.(1983){Baldeschi}, {Gelmini}, and
  {Ruffini}]{baldeschi83}
M.~R. {Baldeschi}, G.~B. {Gelmini}, and R.~{Ruffini}.
\newblock {On massive fermions and bosons in galactic halos}.
\newblock \emph{Physics Letters B}, 122:\penalty0 221--224, March 1983.
\newblock \doi{10.1016/0370-2693(83)90688-3}.

\bibitem[{Preskill} et~al.(1983){Preskill}, {Wise}, and {Wilczek}]{preskill83}
J.~{Preskill}, M.~B. {Wise}, and F.~{Wilczek}.
\newblock {Cosmology of the invisible axion}.
\newblock \emph{Physics Letters B}, 120:\penalty0 127--132, January 1983.
\newblock \doi{10.1016/0370-2693(83)90637-8}.

\bibitem[{Sikivie} and {Yang}(2009)]{sikivie}
P.~{Sikivie} and Q.~{Yang}.
\newblock {Bose-Einstein Condensation of Dark Matter Axions}.
\newblock \emph{Physical Review Letters}, 103\penalty0 (11):\penalty0 111301,
  September 2009.
\newblock \doi{10.1103/PhysRevLett.103.111301}.

\bibitem[{Zhang} and {Chiueh}(2017)]{zhang17}
U.-H. {Zhang} and T.~{Chiueh}.
\newblock {Evolution of linear wave dark matter perturbations in the
  radiation-dominant era}.
\newblock \emph{ArXiv e-prints}, February 2017.

\bibitem[{Sikivie}(1983)]{sikivie83}
P.~{Sikivie}.
\newblock {Experimental tests of the 'invisible' axion}.
\newblock \emph{Physical Review Letters}, 51:\penalty0 1415--1417, October
  1983.
\newblock \doi{10.1103/PhysRevLett.51.1415}.

\bibitem[{Marsh} and {Silk}(2014)]{marsh}
D.~J.~E. {Marsh} and J.~{Silk}.
\newblock {A model for halo formation with axion mixed dark matter}.
\newblock \emph{\mnras}, 437:\penalty0 2652--2663, January 2014.
\newblock \doi{10.1093/mnras/stt2079}.

\bibitem[{Lee} and {Koh}(1996)]{lee96}
J.-W. {Lee} and I.-G. {Koh}.
\newblock {Galactic halos as boson stars}.
\newblock \emph{\prd}, 53:\penalty0 2236--2239, February 1996.
\newblock \doi{10.1103/PhysRevD.53.2236}.

\bibitem[{Guzm{\'a}n} et~al.(1999){Guzm{\'a}n}, {Matos}, and
  {Villegas}]{guzman99}
F.~S. {Guzm{\'a}n}, T.~{Matos}, and H.~B. {Villegas}.
\newblock {Scalar fields as dark matter in spiral galaxies: comparison with
  experiments}.
\newblock \emph{Astronomische Nachrichten}, 320:\penalty0 97, 1999.

\bibitem[{Zhang} et~al.(2016){Zhang}, {Sming Tsai}, {Cheung}, and
  {Chu}]{zhang16}
J.~{Zhang}, Y.-L. {Sming Tsai}, K.~{Cheung}, and M.-C. {Chu}.
\newblock {Ultra-Light Axion Dark Matter and its impacts on dark halo structure
  in $N$-body simulation}.
\newblock \emph{ArXiv e-prints}, November 2016.

\bibitem[{Calabrese} and {Spergel}(2016)]{calabrese}
E.~{Calabrese} and D.~N. {Spergel}.
\newblock {Ultra-light dark matter in ultra-faint dwarf galaxies}.
\newblock \emph{\mnras}, 460:\penalty0 4397--4402, August 2016.
\newblock \doi{10.1093/mnras/stw1256}.

\bibitem[{Schive} et~al.(2014){Schive}, {Chiueh}, and {Broadhurst}]{schive14}
H.-Y. {Schive}, T.~{Chiueh}, and T.~{Broadhurst}.
\newblock {Cosmic structure as the quantum interference of a coherent dark
  wave}.
\newblock \emph{Nature Physics}, 10:\penalty0 496--499, July 2014.
\newblock \doi{10.1038/nphys2996}.

\bibitem[{Menci} et~al.(2017){Menci}, {Merle}, {Totzauer}, {Schneider},
  {Grazian}, {Castellano}, and {Sanchez}]{menci17}
N.~{Menci}, A.~{Merle}, M.~{Totzauer}, A.~{Schneider}, A.~{Grazian},
  M.~{Castellano}, and N.~G. {Sanchez}.
\newblock {Fundamental physics with the Hubble Frontier Fields: constraining
  Dark Matter models with the abundance of extremely faint and distant
  galaxies}.
\newblock \emph{ArXiv e-prints}, January 2017.

\bibitem[{Amendola} and {Barbieri}(2006)]{amendola06}
L.~{Amendola} and R.~{Barbieri}.
\newblock {Dark matter from an ultra-light pseudo-Goldsone-boson}.
\newblock \emph{Physics Letters B}, 642:\penalty0 192--196, November 2006.
\newblock \doi{10.1016/j.physletb.2006.08.069}.

\bibitem[{Kim} and {Marsh}(2016)]{kim16}
J.~E. {Kim} and D.~J.~E. {Marsh}.
\newblock {An ultralight pseudoscalar boson}.
\newblock \emph{\prd}, 93\penalty0 (2):\penalty0 025027, January 2016.
\newblock \doi{10.1103/PhysRevD.93.025027}.

\bibitem[{Marsh}(2016)]{marsh16}
D.~J.~E. {Marsh}.
\newblock {Axion cosmology}.
\newblock \emph{Physics Reports}, 643:\penalty0 1--79, July 2016.
\newblock \doi{10.1016/j.physrep.2016.06.005}.

\bibitem[{Sarkar} et~al.(2017){Sarkar}, {Sethi}, and {Das}]{sarkar17}
A.~{Sarkar}, S.~K. {Sethi}, and S.~{Das}.
\newblock {The effects of the small-scale behaviour of dark matter power
  spectrum on CMB spectral distortion}.
\newblock \emph{ArXiv e-prints}, January 2017.

\bibitem[{Blas} et~al.(2016){Blas}, {Lopez Nacir}, and {Sibiryakov}]{blas16}
D.~{Blas}, D.~{Lopez Nacir}, and S.~{Sibiryakov}.
\newblock {Ultra-Light Dark Matter Resonates with Binary Pulsars}.
\newblock \emph{ArXiv e-prints}, December 2016.

\bibitem[{Khmelnitsky} and {Rubakov}(2014)]{pulsarfuzzy}
A.~{Khmelnitsky} and V.~{Rubakov}.
\newblock {Pulsar timing signal from ultralight scalar dark matter}.
\newblock \emph{\jcap}, 2:\penalty0 019, February 2014.
\newblock \doi{10.1088/1475-7516/2014/02/019}.

\bibitem[{Banik} et~al.(2017){Banik}, {Christopherson}, {Sikivie}, and
  {Todarello}]{banik17}
N.~{Banik}, A.~J. {Christopherson}, P.~{Sikivie}, and E.~M. {Todarello}.
\newblock {New astrophysical bounds on ultralight axionlike particles
  (ULALPs)}.
\newblock \emph{ArXiv e-prints}, January 2017.

\bibitem[{Weinberg} et~al.(2015){Weinberg}, {Bullock}, {Governato}, {Kuzio de
  Naray}, and {Peter}]{weinberg15}
D.~H. {Weinberg}, J.~S. {Bullock}, F.~{Governato}, R.~{Kuzio de Naray}, and
  A.~H.~G. {Peter}.
\newblock {Cold dark matter: Controversies on small scales}.
\newblock \emph{Proceedings of the National Academy of Science}, 112:\penalty0
  12249--12255, October 2015.
\newblock \doi{10.1073/pnas.1308716112}.

\bibitem[{Meiksin}(2009)]{meiksin09}
A.~A. {Meiksin}.
\newblock {The physics of the intergalactic medium}.
\newblock \emph{Reviews of Modern Physics}, 81:\penalty0 1405--1469, October
  2009.
\newblock \doi{10.1103/RevModPhys.81.1405}.

\bibitem[{McQuinn}(2015)]{mcquinn15}
M.~{McQuinn}.
\newblock {The Evolution of the Intergalactic Medium}.
\newblock \emph{ArXiv e-prints}, November 2015.

\bibitem[{Croft} et~al.(2002){Croft}, {Weinberg}, {Bolte}, {Burles},
  {Hernquist}, {Katz}, {Kirkman}, and {Tytler}]{croft02}
R.~A.~C. {Croft}, D.~H. {Weinberg}, M.~{Bolte}, S.~{Burles}, L.~{Hernquist},
  N.~{Katz}, D.~{Kirkman}, and D.~{Tytler}.
\newblock {Toward a Precise Measurement of Matter Clustering: Ly{$\alpha$}
  Forest Data at Redshifts 2-4}.
\newblock \emph{\apj}, 581:\penalty0 20--52, December 2002.
\newblock \doi{10.1086/344099}.

\bibitem[{Zaldarriaga} et~al.(2003){Zaldarriaga}, {Scoccimarro}, and
  {Hui}]{zaldarriaga03}
M.~{Zaldarriaga}, R.~{Scoccimarro}, and L.~{Hui}.
\newblock {Inferring the Linear Power Spectrum from the Ly{$\alpha$} Forest}.
\newblock \emph{\apj}, 590:\penalty0 1--7, June 2003.
\newblock \doi{10.1086/374407}.

\bibitem[{McDonald}(2003)]{mcdonald03}
P.~{McDonald}.
\newblock {Toward a Measurement of the Cosmological Geometry at z \~{} 2:
  Predicting Ly{$\alpha$} Forest Correlation in Three Dimensions and the
  Potential of Future Data Sets}.
\newblock \emph{\apj}, 585:\penalty0 34--51, March 2003.
\newblock \doi{10.1086/345945}.

\bibitem[{Viel} et~al.(2004){Viel}, {Haehnelt}, and {Springel}]{viel04}
M.~{Viel}, M.~G. {Haehnelt}, and V.~{Springel}.
\newblock {Inferring the dark matter power spectrum from the Lyman {$\alpha$}
  forest in high-resolution QSO absorption spectra}.
\newblock \emph{\mnras}, 354:\penalty0 684--694, November 2004.
\newblock \doi{10.1111/j.1365-2966.2004.08224.x}.

\bibitem[{McDonald} et~al.(2005){McDonald}, {Seljak}, {Cen}, {Shih},
  {Weinberg}, {Burles}, {Schneider}, {Schlegel}, {Bahcall}, {Briggs},
  {Brinkmann}, {Fukugita}, {Ivezi{\'c}}, {Kent}, and {Vanden Berk}]{mcdonald05}
P.~{McDonald}, U.~{Seljak}, R.~{Cen}, D.~{Shih}, D.~H. {Weinberg}, S.~{Burles},
  D.~P. {Schneider}, D.~J. {Schlegel}, N.~A. {Bahcall}, J.~W. {Briggs},
  J.~{Brinkmann}, M.~{Fukugita}, {\v Z}.~{Ivezi{\'c}}, S.~{Kent}, and D.~E.
  {Vanden Berk}.
\newblock {The Linear Theory Power Spectrum from the Ly{$\alpha$} Forest in the
  Sloan Digital Sky Survey}.
\newblock \emph{\apj}, 635:\penalty0 761--783, December 2005.
\newblock \doi{10.1086/497563}.

\bibitem[{Seljak} et~al.(2006{\natexlab{a}}){Seljak}, {Slosar}, and
  {McDonald}]{seljak06}
U.~{Seljak}, A.~{Slosar}, and P.~{McDonald}.
\newblock {Cosmological parameters from combining the Lyman-{$\alpha$} forest
  with CMB, galaxy clustering and SN constraints}.
\newblock \emph{\jcap}, 10:\penalty0 014, October 2006{\natexlab{a}}.
\newblock \doi{10.1088/1475-7516/2006/10/014}.

\bibitem[{Lidz} et~al.(2010){Lidz}, {Faucher-Gigu{\`e}re}, {Dall'Aglio},
  {McQuinn}, {Fechner}, {Zaldarriaga}, {Hernquist}, and {Dutta}]{lidz10}
A.~{Lidz}, C.-A. {Faucher-Gigu{\`e}re}, A.~{Dall'Aglio}, M.~{McQuinn},
  C.~{Fechner}, M.~{Zaldarriaga}, L.~{Hernquist}, and S.~{Dutta}.
\newblock {A Measurement of Small-scale Structure in the 2.2-4.2 Ly{$\alpha$}
  Forest}.
\newblock \emph{\apj}, 718:\penalty0 199--230, July 2010.
\newblock \doi{10.1088/0004-637X/718/1/199}.

\bibitem[{Viel} et~al.(2005){Viel}, {Lesgourgues}, {Haehnelt}, {Matarrese}, and
  {Riotto}]{viel05}
M.~{Viel}, J.~{Lesgourgues}, M.~G. {Haehnelt}, S.~{Matarrese}, and A.~{Riotto}.
\newblock {Constraining warm dark matter candidates including sterile neutrinos
  and light gravitinos with WMAP and the Lyman-{$\alpha$} forest}.
\newblock \emph{\prd}, 71\penalty0 (6):\penalty0 063534, March 2005.
\newblock \doi{10.1103/PhysRevD.71.063534}.

\bibitem[{Viel} et~al.(2013){Viel}, {Becker}, {Bolton}, and
  {Haehnelt}]{viel13WDM}
M.~{Viel}, G.~D. {Becker}, J.~S. {Bolton}, and M.~G. {Haehnelt}.
\newblock {Warm dark matter as a solution to the small scale crisis: New
  constraints from high redshift Lyman-{$\alpha$} forest data}.
\newblock \emph{\prd}, 88\penalty0 (4):\penalty0 043502, August 2013.
\newblock \doi{10.1103/PhysRevD.88.043502}.

\bibitem[{Seljak} et~al.(2006{\natexlab{b}}){Seljak}, {Makarov}, {McDonald},
  and {Trac}]{uros06}
U.~{Seljak}, A.~{Makarov}, P.~{McDonald}, and H.~{Trac}.
\newblock {Can Sterile Neutrinos Be the Dark Matter?}
\newblock \emph{Physical Review Letters}, 97\penalty0 (19):\penalty0 191303,
  November 2006{\natexlab{b}}.
\newblock \doi{10.1103/PhysRevLett.97.191303}.

\bibitem[{Baur} et~al.(2016){Baur}, {Palanque-Delabrouille}, {Y{\`e}che},
  {Magneville}, and {Viel}]{baur15}
J.~{Baur}, N.~{Palanque-Delabrouille}, C.~{Y{\`e}che}, C.~{Magneville}, and
  M.~{Viel}.
\newblock {Lyman-alpha forests cool warm dark matter}.
\newblock \emph{\jcap}, 8:\penalty0 012, August 2016.
\newblock \doi{10.1088/1475-7516/2016/08/012}.

\bibitem[{Viel} et~al.(2008){Viel}, {Becker}, {Bolton}, {Haehnelt}, {Rauch},
  and {Sargent}]{viel08}
M.~{Viel}, G.~D. {Becker}, J.~S. {Bolton}, M.~G. {Haehnelt}, M.~{Rauch}, and
  W.~L.~W. {Sargent}.
\newblock {How Cold Is Cold Dark Matter? Small-Scales Constraints from the Flux
  Power Spectrum of the High-Redshift Lyman-{$\alpha$} Forest}.
\newblock \emph{Physical Review Letters}, 100\penalty0 (4):\penalty0 041304,
  February 2008.
\newblock \doi{10.1103/PhysRevLett.100.041304}.

\bibitem[{Yeche} et~al.(2017){Yeche}, {Palanque-Delabrouille}, {.~Baur}, and
  {du Mas des BourBoux}]{yeche17}
C.~{Yeche}, N.~{Palanque-Delabrouille}, J~{.~Baur}, and H.~{du Mas des
  BourBoux}.
\newblock {Constraints on neutrino masses from Lyman-alpha forest power
  spectrum with BOSS and XQ-100}.
\newblock \emph{ArXiv e-prints}, February 2017.

\bibitem[{Palanque-Delabrouille} et~al.(2015){Palanque-Delabrouille},
  {Y{\`e}che}, {Baur}, {Magneville}, {Rossi}, {Lesgourgues}, {Borde}, {Burtin},
  {LeGoff}, {Rich}, {Viel}, and {Weinberg}]{palanque15}
N.~{Palanque-Delabrouille}, C.~{Y{\`e}che}, J.~{Baur}, C.~{Magneville},
  G.~{Rossi}, J.~{Lesgourgues}, A.~{Borde}, E.~{Burtin}, J.-M. {LeGoff},
  J.~{Rich}, M.~{Viel}, and D.~{Weinberg}.
\newblock {Neutrino masses and cosmology with Lyman-alpha forest power
  spectrum}.
\newblock \emph{\jcap}, 11:\penalty0 011, November 2015.
\newblock \doi{10.1088/1475-7516/2015/11/011}.

\bibitem[{Busca} et~al.(2013){Busca}, {Delubac}, {Rich}, {Bailey},
  {Font-Ribera}, {Kirkby}, {Le Goff}, {Pieri}, {Slosar}, {Aubourg}, {Bautista},
  {Bizyaev}, {Blomqvist}, {Bolton}, {Bovy}, {Brewington}, {Borde}, {Brinkmann},
  {Carithers}, {Croft}, {Dawson}, {Ebelke}, {Eisenstein}, {Hamilton}, {Ho},
  {Hogg}, {Honscheid}, {Lee}, {Lundgren}, {Malanushenko}, {Malanushenko},
  {Margala}, {Maraston}, {Mehta}, {Miralda-Escud{\'e}}, {Myers}, {Nichol},
  {Noterdaeme}, {Olmstead}, {Oravetz}, {Palanque-Delabrouille}, {Pan},
  {P{\^a}ris}, {Percival}, {Petitjean}, {Roe}, {Rollinde}, {Ross}, {Rossi},
  {Schlegel}, {Schneider}, {Shelden}, {Sheldon}, {Simmons}, {Snedden},
  {Tinker}, {Viel}, {Weaver}, {Weinberg}, {White}, {Y{\`e}che}, and
  {York}]{busca13}
N.~G. {Busca}, T.~{Delubac}, J.~{Rich}, S.~{Bailey}, A.~{Font-Ribera},
  D.~{Kirkby}, J.-M. {Le Goff}, M.~M. {Pieri}, A.~{Slosar}, {\'E}.~{Aubourg},
  J.~E. {Bautista}, D.~{Bizyaev}, M.~{Blomqvist}, A.~S. {Bolton}, J.~{Bovy},
  H.~{Brewington}, A.~{Borde}, J.~{Brinkmann}, B.~{Carithers}, R.~A.~C.
  {Croft}, K.~S. {Dawson}, G.~{Ebelke}, D.~J. {Eisenstein}, J.-C. {Hamilton},
  S.~{Ho}, D.~W. {Hogg}, K.~{Honscheid}, K.-G. {Lee}, B.~{Lundgren},
  E.~{Malanushenko}, V.~{Malanushenko}, D.~{Margala}, C.~{Maraston},
  K.~{Mehta}, J.~{Miralda-Escud{\'e}}, A.~D. {Myers}, R.~C. {Nichol},
  P.~{Noterdaeme}, M.~D. {Olmstead}, D.~{Oravetz}, N.~{Palanque-Delabrouille},
  K.~{Pan}, I.~{P{\^a}ris}, W.~J. {Percival}, P.~{Petitjean}, N.~A. {Roe},
  E.~{Rollinde}, N.~P. {Ross}, G.~{Rossi}, D.~J. {Schlegel}, D.~P. {Schneider},
  A.~{Shelden}, E.~S. {Sheldon}, A.~{Simmons}, S.~{Snedden}, J.~L. {Tinker},
  M.~{Viel}, B.~A. {Weaver}, D.~H. {Weinberg}, M.~{White}, C.~{Y{\`e}che}, and
  D.~G. {York}.
\newblock {Baryon acoustic oscillations in the Ly{$\alpha$} forest of BOSS
  quasars}.
\newblock \emph{\aap}, 552:\penalty0 A96, April 2013.
\newblock \doi{10.1051/0004-6361/201220724}.

\bibitem[{Slosar} et~al.(2013){Slosar}, {Ir{\v s}i{\v c}}, {Kirkby}, {Bailey},
  {Busca}, {Delubac}, {Rich}, {Aubourg}, {Bautista}, {Bhardwaj}, {Blomqvist},
  {Bolton}, {Bovy}, {Brownstein}, {Carithers}, {Croft}, {Dawson},
  {Font-Ribera}, {Le Goff}, {Ho}, {Honscheid}, {Lee}, {Margala}, {McDonald},
  {Medolin}, {Miralda-Escud{\'e}}, {Myers}, {Nichol}, {Noterdaeme},
  {Palanque-Delabrouille}, {P{\^a}ris}, {Petitjean}, {Pieri}, {Pi{\v s}kur},
  {Roe}, {Ross}, {Rossi}, {Schlegel}, {Schneider}, {Suzuki}, {Sheldon},
  {Seljak}, {Viel}, {Weinberg}, and {Y{\`e}che}]{slosar13}
A.~{Slosar}, V.~{Ir{\v s}i{\v c}}, D.~{Kirkby}, S.~{Bailey}, N.~G. {Busca},
  T.~{Delubac}, J.~{Rich}, {\'E}.~{Aubourg}, J.~E. {Bautista}, V.~{Bhardwaj},
  M.~{Blomqvist}, A.~S. {Bolton}, J.~{Bovy}, J.~{Brownstein}, B.~{Carithers},
  R.~A.~C. {Croft}, K.~S. {Dawson}, A.~{Font-Ribera}, J.-M. {Le Goff}, S.~{Ho},
  K.~{Honscheid}, K.-G. {Lee}, D.~{Margala}, P.~{McDonald}, B.~{Medolin},
  J.~{Miralda-Escud{\'e}}, A.~D. {Myers}, R.~C. {Nichol}, P.~{Noterdaeme},
  N.~{Palanque-Delabrouille}, I.~{P{\^a}ris}, P.~{Petitjean}, M.~M. {Pieri},
  Y.~{Pi{\v s}kur}, N.~A. {Roe}, N.~P. {Ross}, G.~{Rossi}, D.~J. {Schlegel},
  D.~P. {Schneider}, N.~{Suzuki}, E.~S. {Sheldon}, U.~{Seljak}, M.~{Viel},
  D.~H. {Weinberg}, and C.~{Y{\`e}che}.
\newblock {Measurement of baryon acoustic oscillations in the Lyman-{$\alpha$}
  forest fluctuations in BOSS data release 9}.
\newblock \emph{\jcap}, 4:\penalty0 026, April 2013.
\newblock \doi{10.1088/1475-7516/2013/04/026}.

\bibitem[{Garzilli} et~al.(2015){Garzilli}, {Boyarsky}, and
  {Ruchayskiy}]{garzilli15}
A.~{Garzilli}, A.~{Boyarsky}, and O.~{Ruchayskiy}.
\newblock {Cutoff in the Lyman $\{$$\backslash$alpha$\}$ forest power spectrum:
  warm IGM or warm dark matter?}
\newblock \emph{ArXiv e-prints}, October 2015.

\bibitem[{Ir{\v s}i{\v c}} et~al.(2016){Ir{\v s}i{\v c}}, {Viel}, {Berg},
  {D'Odorico}, {Haehnelt}, {Cristiani}, {Cupani}, {Kim}, {L{\'o}pez},
  {Ellison}, {Becker}, {Christensen}, {Denney}, {Worseck}, and
  {Bolton}]{irsic17}
V.~{Ir{\v s}i{\v c}}, M.~{Viel}, T.~A.~M. {Berg}, V.~{D'Odorico}, M.~G.
  {Haehnelt}, S.~{Cristiani}, G.~{Cupani}, T.-S. {Kim}, S.~{L{\'o}pez},
  S.~{Ellison}, G.~D. {Becker}, L.~{Christensen}, K.~D. {Denney}, G.~{Worseck},
  and J.~S. {Bolton}.
\newblock {The Lyman-alpha forest power spectrum from the XQ-100 Legacy
  Survey}.
\newblock \emph{\mnras}, December 2016.
\newblock \doi{10.1093/mnras/stw3372}.

\bibitem[{L{\'o}pez} et~al.(2016){L{\'o}pez}, {D'Odorico}, {Ellison}, {Becker},
  {Christensen}, {Cupani}, {Denney}, {P{\^a}ris}, {Worseck}, {Berg},
  {Cristiani}, {Dessauges-Zavadsky}, {Haehnelt}, {Hamann}, {Hennawi}, {Ir{\v
  s}i{\v c}}, {Kim}, {L{\'o}pez}, {Lund Saust}, {M{\'e}nard}, {Perrotta},
  {Prochaska}, {S{\'a}nchez-Ram{\'{\i}}rez}, {Vestergaard}, {Viel}, and
  {Wisotzki}]{lopez16}
S.~{L{\'o}pez}, V.~{D'Odorico}, S.~L. {Ellison}, G.~D. {Becker},
  L.~{Christensen}, G.~{Cupani}, K.~D. {Denney}, I.~{P{\^a}ris}, G.~{Worseck},
  T.~A.~M. {Berg}, S.~{Cristiani}, M.~{Dessauges-Zavadsky}, M.~{Haehnelt},
  F.~{Hamann}, J.~{Hennawi}, V.~{Ir{\v s}i{\v c}}, T.-S. {Kim}, P.~{L{\'o}pez},
  R.~{Lund Saust}, B.~{M{\'e}nard}, S.~{Perrotta}, J.~X. {Prochaska},
  R.~{S{\'a}nchez-Ram{\'{\i}}rez}, M.~{Vestergaard}, M.~{Viel}, and
  L.~{Wisotzki}.
\newblock {XQ-100: A legacy survey of one hundred $z=3.5-4.5$ quasars observed
  with VLT/X-shooter}.
\newblock \emph{\aap}, 594:\penalty0 A91, October 2016.
\newblock \doi{10.1051/0004-6361/201628161}.

\bibitem[Springel(2005)]{gadget}
Volker Springel.
\newblock {The cosmological simulation code GADGET-2}.
\newblock \emph{\mnras}, 364:\penalty0 1105--1134, 2005.
\newblock \doi{10.1111/j.1365-2966.2005.09655.x}.
\newblock URL \url{http://www.mpa-garching.mpg.de/gadget/}.

\bibitem[{Bolton} et~al.(2016){Bolton}, {Puchwein}, {Sijacki}, {Haehnelt},
  {Kim}, {Meiksin}, {Regan}, and {Viel}]{bolton16}
J.~S. {Bolton}, E.~{Puchwein}, D.~{Sijacki}, M.~G. {Haehnelt}, T.-S. {Kim},
  A.~{Meiksin}, J.~A. {Regan}, and M.~{Viel}.
\newblock {The Sherwood simulation suite: overview and data comparisons with
  the Lyman-alpha forest at redshifts $2 < z < 5$}.
\newblock \emph{ArXiv e-prints}, May 2016.

\bibitem[{Planck Collaboration} et~al.(2016){Planck Collaboration}, {Ade},
  {Aghanim}, {Arnaud}, {Ashdown}, {Aumont}, {Baccigalupi}, {Banday},
  {Barreiro}, {Bartlett}, and et~al.]{planck15}
{Planck Collaboration}, P.~A.~R. {Ade}, N.~{Aghanim}, M.~{Arnaud},
  M.~{Ashdown}, J.~{Aumont}, C.~{Baccigalupi}, A.~J. {Banday}, R.~B.
  {Barreiro}, J.~G. {Bartlett}, and et~al.
\newblock {Planck 2015 results. XIII. Cosmological parameters}.
\newblock \emph{\aap}, 594:\penalty0 A13, September 2016.
\newblock \doi{10.1051/0004-6361/201525830}.

\bibitem[{Schive} et~al.(2016){Schive}, {Chiueh}, {Broadhurst}, and
  {Huang}]{schive16}
H.-Y. {Schive}, T.~{Chiueh}, T.~{Broadhurst}, and K.-W. {Huang}.
\newblock {Contrasting Galaxy Formation from Quantum Wave Dark Matter,
  {$\psi$}DM, with {$\Lambda$}CDM, using Planck and Hubble Data}.
\newblock \emph{\apj}, 818:\penalty0 89, February 2016.
\newblock \doi{10.3847/0004-637X/818/1/89}.

\bibitem[{Bolton} et~al.(2008){Bolton}, {Viel}, {Kim}, {Haehnelt}, and
  {Carswell}]{bolton08}
J.~S. {Bolton}, M.~{Viel}, T.-S. {Kim}, M.~G. {Haehnelt}, and R.~F. {Carswell}.
\newblock {Possible evidence for an inverted temperature-density relation in
  the intergalactic medium from the flux distribution of the Ly{$\alpha$}
  forest}.
\newblock \emph{\mnras}, 386:\penalty0 1131--1144, May 2008.
\newblock \doi{10.1111/j.1365-2966.2008.13114.x}.

\bibitem[{Becker} et~al.(2011){Becker}, {Bolton}, {Haehnelt}, and
  {Sargent}]{becker11}
G.~D. {Becker}, J.~S. {Bolton}, M.~G. {Haehnelt}, and W.~L.~W. {Sargent}.
\newblock {Detection of extended He II reionization in the temperature
  evolution of the intergalactic medium}.
\newblock \emph{\mnras}, 410:\penalty0 1096--1112, January 2011.
\newblock \doi{10.1111/j.1365-2966.2010.17507.x}.

\bibitem[sup()]{supplemental}
See Supplemental Material at [url] for a discussion on how the temperature and
  redshift of reionization affect the thermal Jeans smoothing, which includes
  \citep{gnedin98,kulkarni15}.

\bibitem[{Palanque-Delabrouille} et~al.(2013){Palanque-Delabrouille},
  {Y{\`e}che}, {Borde}, {Le Goff}, {Rossi}, {Viel}, {Aubourg}, {Bailey},
  {Bautista}, {Blomqvist}, {Bolton}, {Bolton}, {Busca}, {Carithers}, {Croft},
  {Dawson}, {Delubac}, {Font-Ribera}, {Ho}, {Kirkby}, {Lee}, {Margala},
  {Miralda-Escud{\'e}}, {Muna}, {Myers}, {Noterdaeme}, {P{\^a}ris},
  {Petitjean}, {Pieri}, {Rich}, {Rollinde}, {Ross}, {Schlegel}, {Schneider},
  {Slosar}, and {Weinberg}]{palanque13}
N.~{Palanque-Delabrouille}, C.~{Y{\`e}che}, A.~{Borde}, J.-M. {Le Goff},
  G.~{Rossi}, M.~{Viel}, {\'E}.~{Aubourg}, S.~{Bailey}, J.~{Bautista},
  M.~{Blomqvist}, A.~{Bolton}, J.~S. {Bolton}, N.~G. {Busca}, B.~{Carithers},
  R.~A.~C. {Croft}, K.~S. {Dawson}, T.~{Delubac}, A.~{Font-Ribera}, S.~{Ho},
  D.~{Kirkby}, K.-G. {Lee}, D.~{Margala}, J.~{Miralda-Escud{\'e}}, D.~{Muna},
  A.~D. {Myers}, P.~{Noterdaeme}, I.~{P{\^a}ris}, P.~{Petitjean}, M.~M.
  {Pieri}, J.~{Rich}, E.~{Rollinde}, N.~P. {Ross}, D.~J. {Schlegel}, D.~P.
  {Schneider}, A.~{Slosar}, and D.~H. {Weinberg}.
\newblock {The one-dimensional Ly{$\alpha$} forest power spectrum from BOSS}.
\newblock \emph{\aap}, 559:\penalty0 A85, November 2013.
\newblock \doi{10.1051/0004-6361/201322130}.

\bibitem[{Puchwein} et~al.(2015){Puchwein}, {Bolton}, {Haehnelt}, {Madau},
  {Becker}, and {Haardt}]{puchwein15}
E.~{Puchwein}, J.~S. {Bolton}, M.~G. {Haehnelt}, P.~{Madau}, G.~D. {Becker},
  and F.~{Haardt}.
\newblock {The photoheating of the intergalactic medium in synthesis models of
  the UV background}.
\newblock \emph{\mnras}, 450:\penalty0 4081--4097, July 2015.
\newblock \doi{10.1093/mnras/stv773}.

\bibitem[{Upton Sanderbeck} et~al.(2016){Upton Sanderbeck}, {D'Aloisio}, and
  {McQuinn}]{upton}
P.~R. {Upton Sanderbeck}, A.~{D'Aloisio}, and M.~J. {McQuinn}.
\newblock {Models of the thermal evolution of the intergalactic medium after
  reionization}.
\newblock \emph{\mnras}, 460:\penalty0 1885--1897, August 2016.
\newblock \doi{10.1093/mnras/stw1117}.

\bibitem[{Armengaud} et~al.(2017){Armengaud}, {Palanque-Delabrouille},
  {Y{\`e}che}, {Marsh}, and {Baur}]{armengaud17}
E.~{Armengaud}, N.~{Palanque-Delabrouille}, C.~{Y{\`e}che}, D.~J.~E. {Marsh},
  and J.~{Baur}.
\newblock {Constraining the mass of light bosonic dark matter using SDSS
  Lyman-$\alpha$ forest}.
\newblock \emph{ArXiv e-prints}, March 2017.

\bibitem[Gnedin and Hui(1998)]{gnedin98}
Nickolay~Y. Gnedin and Lam Hui.
\newblock {Probing the universe with the Lyman alpha forest: 1. Hydrodynamics
  of the low density IGM}.
\newblock \emph{Mon. Not. Roy. Astron. Soc.}, 296:\penalty0 44--55, 1998.
\newblock \doi{10.1046/j.1365-8711.1998.01249.x}.

\bibitem[{Kulkarni} et~al.(2015){Kulkarni}, {Hennawi}, {O{\~n}orbe}, {Rorai},
  and {Springel}]{kulkarni15}
G.~{Kulkarni}, J.~F. {Hennawi}, J.~{O{\~n}orbe}, A.~{Rorai}, and V.~{Springel}.
\newblock {Characterizing the Pressure Smoothing Scale of the Intergalactic
  Medium}.
\newblock \emph{\apj}, 812:\penalty0 30, October 2015.
\newblock \doi{10.1088/0004-637X/812/1/30}.

\end{thebibliography}
